\documentstyle[eqsecnum,aps]{revtex}

\begin{document}
\draft
\title{Polarizable particles aggregation under rotating magnetic fields
using scattering dichroism}
\author{Sonia Melle$^{1,2}$, Oscar G. Calder\'on$^{3}$,
Gerald G. Fuller$^{1}$, Miguel A. Rubio$^{2}$}

\address{$^{1}$ Dpt. Chemical Engineering, Stanford University,
Stanford, CA 94305-5025, USA  }
\address{$^{2}$ Dpto. F\'{\i}sica Fundamental, UNED. Paseo Senda del Rey 9, Madrid
28040, Spain }
\address{$^{3}$ Dpto. Optica, Universidad Complutense de Madrid, Ciudad
Universitaria, Madrid 28040, Spain}

\date{\today}
\maketitle

\begin{abstract}
We used scattering dichroism to study the combined effects of
viscous and magnetic forces on the dynamics of dipolar chains
induced in magnetorheological suspensions under rotating magnetic
fields. We found that the chains adjust their size to rotate
synchronously with the field but with a constant phase lag. Two
different behaviors for the dichroism (proportional to the total
number of aggregated particles) and the phase lag are found below
or above a critical frequency. We obtained a linear dependence of
the critical frequency with the square of the magnetization and
with the inverse of the viscosity. The Mason number (ratio of
viscous to magnetic forces) governs the dynamics. Therefore there
is a critical Mason number below which, the dichroism remains
almost constant and above which, the rotation of the field
prevents the particle aggregation process from taken place being
this the mechanism responsible for the decrease of dichroism. Our
experimental results have been corroborated with particle dynamics
simulations showing good agreement.
\end{abstract}

\pacs{PACS numbers:  }
\preprint{HEP/123-qed}
\widetext

\section{Introduction}
Magnetorheological (MR) suspensions consist of magnetizable
particles suspended in a nonmagnetic fluid. They are a model
system for the study of structure formation and dynamics in
dipolar suspensions with tunable interaction between the
particles. It is well known that when a unidirectional magnetic
field is applied, the particles making up the suspension acquire a
dipole moment $\vec{m}=(4\pi /3)a^{3}\vec{M}$, where $\vec{M}$ and
$a$ are the particle magnetization and diameter, respectively. Due
to dipolar interactions, these particles join to form chain
structures in the field direction inducing an optical anisotropy
in the sample due to polarization dependent scattering from
oriented aggregates. When the size of the scattering objects is
comparable to the wavelength of the light ($\lambda_l$), the
induced dichroism will be bigger than the induced birefringence
\cite{note1,Vandehulst81}. Due to the polarization scattering,
scattering dichroism has been shown to be a good technique for the
study of these anisotropic structures in MR suspensions at
moderate concentration which cannot be readily studied with other
optical techniques, such as video microscopy. Using Rayleigh-Debye
theory for the scattering of light we can estimate the scattering
dichroism generated from a chain $j$ formed by $N_j$ particles in
the forward direction ($\phi=0$) to be \cite{Vandehulst81}:

\begin{equation} \label{RAYLEIGH}
\Delta n^{\prime\prime}_j = \frac {2 N_j}{k^2} \left( Re \left[
T_2(\phi=0)\right] - Re \left[ T_1(\phi=0)\right]\right) \; ,
\end{equation}

\noindent where $k = 2 \pi / \lambda_l $ is the wave vector of the
laser beam and $ T_i(\phi=0) $ is a function of $k, a,$ and the
isotropic refractive indexes of solvent and particles. We assume
that the scattering dichroism produced from a set of $N_c$ chains
is the incoherent sum of the scattering dichroism produced for
each chain, so the total dichroism is proportional to the total
number of aggregated particles $N_{a}$:

\begin{equation} \label{DICHROISM}
\Delta n^{\prime\prime} = \sum^{N_c}_{j=1} \Delta
n^{\prime\prime}_j \propto \sum^{N_c}_{j=1} N_j \equiv N_{a} \; .
\end{equation}

Recently, we studied the dynamics of MR suspensions under rotating
magnetic fields \cite{Melle00}. We found that the field-induced
chains rotate synchronously with the field but lag behind by a
constant phase angle. Previous experimental work on magnetic holes
\cite{Helgesen90} and liquid crystals \cite{Meyer91,Meyer97} under
rotating magnetic fields report synchronous and non-synchronous
regimes depending on the value of the driving frequency. However,
as we showed previously \cite{Melle00}, MR suspensions have the
capacity of reducing the size of the structures to decrease their
viscous drag while rotating synchronously with the field
\cite{note2}. Furthermore, within this synchronous regime, two
different behaviors were found below or above a critical
frequency. Below this value, the dichroism remains almost constant
but above the critical frequency the viscous drag overcomes the
magnetic force and reduces the dichroism following a power law
with an exponent -1. The critical frequency was found to increase
with the amplitude of the applied field.

In this paper we analyze the interplay between magnetic and
viscous forces over the critical frequency separating these two
regimes. We have studied the dependence of the critical frequency
on the magnetization by applying magnetic fields with different
amplitudes on the same suspension. The dependence of the critical
frequency on the viscosity of the carrier fluid was analyzed
applying a constant field on suspensions with different glycerol
concentrations.

A dimensionless parameter that compares these two forces is the
well-known Mason number (ratio of viscous to magnetic forces).
This number has been defined with different proportionality
factors in literature \cite{Gast89,Bossis99,Martin00}. The Stokes
viscous force acting on two particles in contact which rotate with
a field of frequency $\omega$  is $F_h \sim 6 \pi \eta a (\omega
a)$, where $\eta$ is the solvent viscosity. The dipolar magnetic
force is $F_m \sim (\pi/2) \mu_0 a^2 M^2 $, where $\mu _{0}$ is
the vacuum magnetic permeability. Then, the Mason number is given
by:

\begin{equation} \label{MASON}
M\!a=\frac{12^{2}\eta \omega}{\mu _{0}M^{2}}\;,
\end{equation}

\noindent where the proportionality factor was chosen to be in
agreement with the dimensionless frequency obtained from the
theoretical analysis (Section \ref{simulations}).

We found that the critical frequency separating the two regimes
increases linearly with the square of the magnetization and
decreases with the inverse of the viscosity, so the Mason number
governs the dynamics of field-induced dipolar chains under
rotating fields. As expected from this result, we obtained a good
collapse of the dichroism and the phase lag curves (measured at
different magnetic fields and viscosities) with Mason number. The
change in behavior of the dichroism and the phase lag occurs at a
critical Mason number, $M\!a_c \approx 1$, above which the viscous
forces dominate and inhibit the aggregation process. We
corroborated our experimental findings through the results
obtained from particle dynamics simulations. The simulations show
that the average length of the chains decreases with frequency
following a power law with an exponent close to - 0.5.
Furthermore, the simulations also reveal the two different
behaviors for the total number of aggregated particles and the
phase lag and are in good agreement with the experiments.

\section{Materials and Methods}

\subsection{MR suspensions and procedure}

We prepared glycerol suspensions at different concentrations using
two water suspensions of polystyrene (PS) polydisperse
micro-spheres loaded with iron oxide grains. The acqueous
suspensions (M1-180/12 and M1-070/60) were acquired commercially
from {\it Estapor-$Rh\hat{o}ne \;Poulenc$} with a solid content of
10$\%$. The particle properties are list in Table \ref{Table1}.

The surface of these microspheres is composed of carboxylic acid
(-COOH) groups with an added surfactant coating layer of sodium
dodecyl sulfate (SDS) to stabilize the suspensions. Since these
small magnetite grains are randomly oriented inside the
micro-particles, the resulting magnetic moment is zero in the
absence of an external magnetic field. Under sufficiently low
magnetic fields these particles exhibit superparamagnetic behavior
with virtually no hysteresis or magnetic remanence as a result of
the orientation of the magnetite grains dispersed in the PS
matrix.

Two experimental protocols were used and are summarized in Table
\ref{Table2}. To study the effect of the magnetic forces on the
critical frequency, we diluted the suspension of particles of
diameter 1.01 $\mu$m (M1-180/12) in glycerol to achieve a solvent
volume concentration of 82.5$\%$ glycerol with a volume fraction
of $\phi_v = 0.016 $. We will denote the resulting suspension as
d-M1-180/12 hereafter. The dilution in glycerol reduced the SDS
concentration and additional SDS was added to achieve a
concentration equal to the original solution ($5 gr / liter $).
The viscosity of the suspension d-M1-10/12 was measured without
applying a magnetic field using a Rheometrics Dynamic Analyzer
{\em \ RDA II} to be $\eta = (0.178 \pm 0.002)$ Pa s at $10^{0}$
C. This set of experiments consists on applying magnetic fields
with different amplitudes (from $H$ = 6.2 kA/m to $H$ = 24.8 kA/m)
on suspension d-M1-180/12. For high fields, the dipole moment
induced on the particles is not directly proportional to the
applied field. To determine this non-linear response, we measured
the particle magnetization curve with a vibrating sample
magnetometer (VSM)-{\em \ LakeShore 7300} for the range of
magnetic fields used in the experiments. The result is shown in
Figure \ref{f:VSMcurve}. As expected, the particles show a
superparamagnetic behavior with no hysteresis but saturation
effects in the magnetization for high fields appeared.

For the second set of experiments we prepared suspensions with
volume fraction  $\phi_v=0.01$ but different glycerol
concentrations ranging from 40$\%$ to 80$\%$ in volume. In this
case, we used the water suspension M1-070/60 with particles of
diameter 0.90 $\mu$m. These suspensions are denoted as
d-M1-070/60. We also determined their viscosities with the {\em \
RDA II}. The effect of the viscous force on the critical frequency
was studied by applying a magnetic field of $H$ = 3.1 kA/m on the
suspensions d-M1-070/60. For this low magnetic field the
magnetization is linear with the field $\vec{M} $= $\chi_{eff}
\vec{H}$, being the effective magnetic susceptibility of the
particle equal to
 $\chi_{eff} = 3(\mu_p - \mu_s) / (\mu_p + 2 \mu_s)$ = 1.90.
Here $\mu_p$ and $\mu_s$ are the permeability of the particles and
the solvent, respectively \cite{Jones95}.

\subsection{Experimental setup}

To study the dynamics of moderately concentrated magnetic dipolar
suspensions (in the range of 1-2 vol\%) under rotating magnetic
fields scattering dichroism was used. A full description of this
experimental technique can be found in Refs.
\cite{Melle00,Fuller95}. A schematic diagram of the optical train
used to measure linear dichroism is shown in Figure
\ref{f:SETUP}(a) and consists on a He-Ne laser, a polarizer
($0^{0}) $, a photoelastic modulator ($45^{0} $) and a quarter
wave plate ($0^{0} $). The light is then passed through the
sample. The transmitted light, detected by a photodiode, is sent
to two phase lock-in amplifiers and then digitized using a 16-bit
A/D data acquisition device ({\em National Instruments}). With
this optical train we can simultaneously measured the time
evolution of the dichroism $\Delta n^{\prime\prime}
=n_1^{\prime\prime} - n_2^{\prime\prime}$, i.e., the difference
between the absorption of the incident light in the parallel than
in the perpendicular direction to the long axis of the aggregates;
and the orientation angle $\theta^{\prime\prime}$, i.e., the angle
between the reference axis of the optical train and the long axis
of the aggregates (see Fig. \ref{f:SETUP}(b)). By comparing
$\theta^{\prime\prime}$ with the temporal evolution of the
magnetic field direction given by $\omega t $ we found that the
structures follow the magnetic field by rotating with the same
frequency but with a phase lag that is independent on time for all
frequencies measured. Therefore we define the phase lag between
the field and the aggregates as $\alpha (t) = \omega t -
\theta^{\prime\prime} (t)$.

The rotating magnetic field was achieved by applying sinusoidal
electric signals to two orthogonal pairs of coils by means of two
{\em Kepco BOP20-10M} power amplifiers, driven by two {\em \
HP-FG3325A} function generators referenced to one another at a
phase difference of $90^{0} $. In Figure \ref{f:SETUP}(c) we show
a sketch of the coils system from the direction of the laser beam.
The function generators allowed for control of both the amplitude
and the frequency of the rotating magnetic field. These coils are
housed in a temperature controlled aluminum cylinder to prevent
heating effects. The suspension is placed between two quartz
windows held in place by a Delrin attachment with thickness $e =
100 \mu$m (see detail of the sample cell on Fig.
\ref{f:SETUP}(d)). All experiments were performed at a temperature
of $T = 282 \pm 1$ K measured on the sample.

\section{Results and discussion}

\subsection{Experimental results}

When a rotating field in the plane (X,Y) perpendicular to the
optical path direction (axis Z):
\begin{equation} \label{FIELD}
\vec{H}= H \left( \cos(\omega t) \hat{x} + \sin( \omega t )
\hat{y} \right) \; ,
\end{equation}
\noindent is applied, individual particle chains are observed to
continually aggregate and fragment, although after a transient the
ensemble achieves a steady state distribution. In the following
figures we plot the dichroism and phase lag reached at the steady
state, denoted as $ \Delta n^{\prime\prime}_0$ and $\alpha_0$,
respectively.

\subsubsection{Magnetic field effect}

In Figure \ref{f:H_variable_DIC}(a) the dichroism generated by the
suspension d-M1-180/12 (82.5\% glycerol, $\phi_v$ = 0.016) is
plotted versus the magnetic field frequency $(f = \omega / 2\pi )$
in a log-log form for various magnetic field strengths ($H$ = 6.2
- 24.8 kA/m). As the frequency of the applied field is increased,
the dichroism is reduced but not with the same rate. This plot
clearly shows two distinct regions for frequencies below and above
a critical frequency, $f_c$. Below this critical frequency, the
dichroism is essentially independent of frequency and the average
number of particles stays almost constant. However, once this
frequency is surpassed, the dichroism strongly decreases with
frequency, which reveals a diminution of the number of aggregated
particles. It is found above 1 Hz that the dichroism drops with
frequency with a scaling of approximately $\Delta n^{\prime
\prime}_0 \simeq f^{-1}$. For high fields, the change in behavior
is moved to higher frequencies, i.e., the critical frequency
separating these two regions is shown to increase with the
magnitude of the applied magnetic field. At higher fields the
particle chains are more able to rotate with higher frequency
fields. Since the strength of the interparticle magnetic forces
scales with the square of the particle magnetization that
dependence is also expected for the critical frequency. To test
this dependence, $f_c \sim M^2$, we plot the dichroism curves
obtained for different magnetic field strengths versus $M\!a$
number in Figure \ref{f:H_variable_DIC}(b) and observe that the
curves collapse onto a master curve.

The phase difference between the orientation of the aggregates and
the magnetic field, $\alpha_0$, versus the frequency of the
applied field is plotted in Figure \ref{f:H_variable_PHASE}(a) for
different field strengths. We observe that this phase lag
increases with frequency over the whole range of frequencies.
However, as we found for the dichroism results, two different
responses are observed, depending on the magnitude of the
frequency. At low frequencies (below $f_c$) the phase difference
grows very quickly while at high frequencies the increase of the
phase lag is relatively slow. For frequencies larger than the
critical frequency, the chains start to disappear so the
contribution to the total average phase lag is due to the few
small chains that still remain. These small chains lag the
magnetic field with larger phase angles since they are close to
breaking apart. In addition, a diminution of the phase lag with
magnetic field intensity is observed. This behavior is expected
since the angular component of the magnetic force, which is the
responsible for the rotation of the chain, increases with the
amplitude of the magnetic field. These data are replotted versus
$M\!a$ number in Figure \ref{f:H_variable_PHASE}(b) and this also
results in a single master curve.

\subsubsection{Influence of Viscosity}

To analyze the influence of the viscosity of the suspending
medium, we measured the dichroism induced in the suspensions
d-M1-07/60 (with different glycerol concentrations and the same
particle volume fraction, $\phi_v$ = 0.01) when a field of
amplitude $H$ = 3.1 kA/m is applied. In Figure
\ref{f:visc_variable_DIC}(a) we plot the variation of the
dichroism with rotational frequency in a log-log form for
solutions with glycerol content ranging from 40\% to 80\%. Again,
two different regimes above or below a critical frequency are
observed. As expected, the critical frequency moves to bigger
frequencies for low viscosity suspensions because the structures
are more free to rotate synchronously with the field. The collapse
of those curves with Mason number is presented in Figure
\ref{f:visc_variable_DIC}(b) verifying the dependence of the
critical frequency with the inverse of the viscosity, $f_c \sim
\eta^{-1}$. Note that for low frequencies, $f < f_c$, the
dichroism shows a flatter plateau for solutions with less glycerol
content (40-50\%) which may be due to the fact that more viscous
solutions need more time to reach the steady state.

The steady phase lag measured for these suspensions is plotted in
Figure \ref{f:visc_variable_PHASE}(a) versus rotational frequency.
We observe that the phase lag variation with the viscosity is much
larger than the variation encountered with field strength. As
expected, more viscous suspensions show bigger phase lags. The
collapse of those curves with Mason number is plotted in Figure
\ref{f:visc_variable_PHASE}(b).

These results emphasize the importance of the Mason number in
controlling the aggregation phenomena in suspensions of
polarizable particles in rotating magnetic fields. The change in
behavior for both the dichroism and the phase lag corresponds to
$M\!a_c \approx 1$ when the viscous forces begin to overcome the
magnetic forces.

\subsection{Numerical Simulations} \label{simulations}

\subsubsection{Equation of motion}

Particle dynamics simulations were conducted that consider the 2D
aggregation kinetics of $N$ dipolar particles of diameter $2a$
suspended in a fluid of viscosity $\eta$ and subjected to rotating
magnetic fields of amplitude $H$ and angular frequency $\omega$
(see Eq. [\ref{FIELD}]). Two fundamental length scales
characterize the formation of chains. The first one is the
thermo-magnetic equilibrium distance $R_{1}=2a\lambda ^{1/3}$,
where $\lambda$ is the well-known dimensionless parameter
calculated as the ratio between magnetic and thermal energies:
\begin{equation}  \label{LAMBDA}
\lambda = \frac{\pi \mu_0 a^{3} M^2} {9 k_B T} \; ,
\end{equation}
\noindent where $\mu _{0}$ is the vacuum magnetic permeability,
and $k_{B}$ is the Boltzmann constant. The second length scale is
the average initial interparticle distance, which can be estimated
as $ R_{0} \simeq 2a/\phi_{v}^{1/3}$. For the experiments
presented above $R_{1}>R_{0}$ \cite{note3}, so the application of
an external field $H$ will immediately trigger the magnetic
aggregation process. Recognizing this rapid aggregation response,
we can neglect the effect of the Brownian motion on the evolution
of the structures. Therefore the equation of motion of the $i$th
particle will be the sum of the following three forces
\cite{Martin98}:

\begin{equation}  \label{MOTION}
{\rm m} \frac{d \vec{v}_i}{d t} = \vec{F}_h\left(\vec{v}_i\right)
+ \sum_{j \ne i} \vec{F}_d\left(\vec{r%
}_{ij}\right) +\sum_{j \ne i} \vec{F}_{hs}\left(r_{ij}\right) \; ,
\end{equation}

\noindent where $\vec{F}_h\left(\vec{v}_i\right)=-6 \pi \eta a
\vec{v}_i$ is the hydrodynamic Stokes force and $\vec{v}_i$ is the
particle velocity. The dipolar force over particle $i$th will be
the sum of the dipole-dipole forces exerted by all the other
particles over it. The dipole-dipole force between particles $i$th
and $j$th is given by

\begin{equation}  \label{DIPOLAR}
\vec{F}_d(\vec{r}_{ij}) = \frac{3 \mu_0 m^2}{4 \pi {r}^4_{ij}}
\left\{ \left[1-5(\hat{m} \cdot \hat{r}_{ij})^2 \right]
\hat{r}_{ij} + 2 (\hat{m} \cdot \hat{r}_{ij}) \hat{m} \right\} \;
\end{equation}

\noindent where $\vec{r}_{ij}$ is the separation vector between
the two center of mass, and we take $\vec{m}$ to be aligned with
the field direction. $\vec{F}_{hs}$ is a strong repulsive force
between the particles to simulate a hard-sphere interaction
between them when in contact. This force is calculated from
\cite{Mohebi96}

\begin{equation}
\vec{F}_{hs}(r_{ij}) = A \frac{3 \mu_0 m^2}{4 \pi (2 a)^4} \exp
\left[ - B (r_{ij}/(2a) -1) \right] \hat{r}_{ij} \; ,
\end{equation}

\noindent where we assume $A=2$ and $B=10$. We can normally
neglect the inertial term in Eq. [\ref{MOTION}] because the
viscous drag term dominates. In the absence of Brownian motion,
the strength of the dipolar interactions only influences the
coarsening time scale and not the structural evolution
\cite{Martin98}. We take the dimensionless length $\rho= r /(2a)$
and the dimensionless time $\tau = t / \beta$ where $\beta = 12^2
\eta / (\mu_0 M^2)$. This temporal scale leads to a dimensionless
rotating frequency equal to $\Omega \equiv \omega \beta$. This
dimensionless frequency is the well-known Mason number with the
definition used in Eq. [\ref{MASON}], $\Omega \equiv M\!a$.

\subsubsection{Numerical results}

We have simulated the first set of experiments using parameters
that correspond with suspension d-M1-180/12. The volume fraction
$\phi_v=0.016$ corresponds to an initial average separation
between particles equal to $ R_{0} /(2a) \simeq 3.55.$

For each magnetic frequency, $\Omega$, we calculate the
experimental observables by averaging their values during the last
period of the magnetic rotation after steady state had been
reached. As we pointed out previously, $N_{a}$ gives us
information on the dichroism. In order to calculate the total
phase lag, $\alpha$, we average the phase lag of each cluster,
$\alpha_j$, using the number of particles in each cluster, $N_j$
as a weighting function.

\begin{equation}
\alpha =\frac{\sum_{j}N_{j}\alpha _{j}}{\sum_{j}N_{j}}\;,
\label{LAG}
\end{equation}

\noindent where the sum is done for $j$ with $N_{j}>1$. The phase
lag of each cluster is the difference between the orientation of
the magnetic field and the long axis of the chain. To compute the
chain orientation we calculate their principal moments of inertia,
$I_{j}^{max}$, $I_{j}^{min}$. Then, the direction of $
I_{j}^{min}$ gives the orientation of the long axis of the
cluster. We also have examined the behavior of the average length,
$L$

\begin{equation}
L=\frac{\sum_{j}N_{j}}{\sum_{j}1}\;.  \label{LENGTH}
\end{equation}

To compute these physical quantities we considered straight chains
without taking into account the shape of the clusters. To check
its effect in our simulations results, we calculated the same
quantities but using a weight function $W_{j}=N_{j}s_{j}$, where
$s_{j}$ is a shape factor with value 1 for the case of a straight
chain, and 0 for a symmetric cluster

\begin{equation}
s_j= \frac{\left(I_j^{max}\right)^{1/2} -
\left(I_j^{min}\right)^{1/2} } {\left(I_j^{max}\right)^{1/2} +
\left(I_j^{min}\right)^{1/2}} \; .
\end{equation}

We did not observe appreciable changes in the simulations results
using this shape correction. Therefore, we present here the
results corresponding to the case without the shape factor.

In Figure \ref{f:MAP} we plot the particle positions in the (X,Y)
plane for different dimensionless frequencies at one arbitrary
time. As we can see in this figure, the size of the structures
becomes smaller as the rotational frequency increases. The
dimensionless average length versus dimensionless frequency is
shown in Figure \ref{f:SIMULATION_RESULTS}(a). This average length
follows a power law behavior with an exponent equal to -0.45. This
value is very close to the value (-0.5) predicted by the chain
model developed by Martin et al. \cite{Martin96,private} for
electrorheological fluids subject to steady shear. Figure
\ref{f:SIMULATION_RESULTS}(b) shows the number of aggregated
particles versus dimensionless frequency $\Omega$. Two different
regimes appears, in agreement with the experimental dichroism. A
flat shape for low frequencies, and a strongly decreasing response
at frequencies when $\Omega _{c}\simeq 1$ ($M\!a_{c}\simeq 1$) is
surpassed. That means that although the average length is
decreasing monotonically with $M\!a$, this diminution in length
starts affecting the dichroism when the viscous forces dominates
over the magnetic forces, i.e., for average chain lengths close to
two particles. This is the length associated to the critical Mason
number $M\!a_{c}\simeq 1$ (see Fig.
\ref{f:SIMULATION_RESULTS}(b)). For $M\!a > 1$ the $N_{a}$ follows
a power law with an exponent close to $-1$, the same value found
in the experiments.

These results point out that the number of aggregated particles
tracks the dichroism. Therefore we suppose that the theoretical
dichroism, computed by the simulations, is a linear function of
the $N_{a}$. In Figure \ref{f:H_variable_vsMA_simu}(a) the
measured dichroism (markers) and the simulated dichroism (solid
line) are plotted and show good agreement. We compare in Figure
\ref{f:H_variable_vsMA_simu}(b) the simulated phase lag (see Eq.
[\ref{LAG}]) with the experimental values versus dimensionless
frequency $\Omega \equiv M\!a$. As we can see the theoretical
behavior agrees with the experimental results. Below a critical
dimensionless frequency $\Omega _{c}$, the phase lag increases
very quickly and above this value the phase lag remains almost
constant.

\section{Conclusions}

In this work we used scattering dichroism to analyze the influence
of the magnetic and viscous forces on the dynamics of
magnetorheological suspensions under rotating magnetic fields. In
these suspensions, dichroism arises from the formation of
optically anisotropic chains upon imposition of the magnetic field
and measure the number of aggregated particles. We found that
under rotating fields, these chains adjust their size to rotate
synchronously with the field but with a constant phase lag. Two
different behaviors for the dichroism and the phase lag are found
above or below a critical frequency. We obtained a linear
dependence of the critical frequency with the square of the
magnetization and with the inverse of the viscosity. This means
that the Mason number (ratio of viscous to magnetic forces)
governs the dynamics correlating the magnetization and viscosity
dependencies of the critical frequency. Therefore there is a
critical Mason number below which, the dichroism remains almost
constant. Above $M\!a_c$, the rotation of the field prevents the
particle aggregation process from taking place and hydrodynamic
friction forces cause a decrease of dichroism.

Simulations incorporating Stokes friction and dipole-dipole
magnetic interaction are able to reproduce the experimental
results. According to the simulations, the average length of the
chain-like structures decreases when increasing the rotational
frequency, scaling as the inverse square root of the Mason number.
Furthermore, the simulations reproduce two different behaviors
above and below the critical Mason number, $M\!a _{c}$, for the
total number of aggregated particles, $N_a$, and the phase lag,
$\alpha$, in agreement with the experiments.
\acknowledgments

This work was partially supported by the Lai Family Grant. S.M.
was supported by the DGICYT (Grant No. PB96-0148). We gratefully
acknowledge A. Gast for her advice and fruitful discussions.

\begin{table}[tbp]
\caption{ Properties of the magnetic latex micro-spheres used.}
\label{Table1}
\end{table}

\begin{table}[tbp]
\caption{ Summary of the experimental conditions.} \label{Table2}
\end{table}

\begin{figure}[tbp]
\caption{Measured magnetization curve for the particles M1-180/12
in the range of magnetic fields reported in the experiments. The
particles show a superparamagnetic behavior with no hysteresis but
with saturation effects for high fields. } \label{f:VSMcurve}
\end{figure}

\begin{figure}[lp]
\caption{ Experimental setup. (a) Schematic diagram of linear
dichroism optical train which enables one to measure the
dichroism, $\Delta n^{\prime\prime}=n_1^{\prime\prime} -
n_2^{\prime\prime}$, and the orientation angle of the structures,
$\theta^{\prime\prime} $. (b) Definition of the phase lag, $\alpha
(t) = \omega t - \theta^{\prime\prime} (t)$ , between the magnetic
field and the chains. (c) Sketch of the coils system to generate
the rotating magnetic field. (d) Detail of the quartz cell filled
with MR suspension. } \label{f:SETUP}
\end{figure}

\begin{figure}[tbp]
\caption{ Variation of the steady dichroism with rotational
frequency (a) and collapse of dichroism with $M\!a$ number (b) for
different magnetic field strengths in a log-log plot. Measurements
for suspension d-M1-180/12 (82.5$\%$ glycerol content, $\phi_v$ =
0.016). Power law fit with an exponent -1 for $M\!a > Ma_c$. }
\label{f:H_variable_DIC}
\end{figure}

\begin{figure}[tbp]
\vskip0.1in \caption{ Variation of the steady phase lag with
rotational frequency (a) and collapse of the phase lag with $M\!a$
number (b) for different magnetic field strengths. Measurements
for suspension d-M1-180/12 (82.5$\%$ glycerol content, $\phi_v$ =
0.016).} \label{f:H_variable_PHASE}
\end{figure}

\begin{figure}[tbp]
\vskip0.1in \caption{ Variation of the steady dichroism with
rotational frequency (a) and collapse of dichroism with $M\!a$
number (b) for suspensions d-M1-070/60 (different glycerol
content, $\phi_v$ = 0.01) in a log-log plot when a magnetic field
of amplitude $H$= 3.1 kA/m is applied.}
\label{f:visc_variable_DIC}
\end{figure}

\begin{figure}[tbp]
\caption{  Variation of the steady phase lag with rotational
frequency (a) and collapse of the phase lag with $M\!a$ number (b)
for suspensions d-M1-070/60 (different glycerol content, $\phi_v$
= 0.01) in a log-log plot when a magnetic field of amplitude $H$=
3.1 kA/m is applied.} \label{f:visc_variable_PHASE}
\end{figure}

\begin{figure}[tbp]
\vskip0.1in \caption{ Dimensionless particles position for
different dimensionless rotating frequencies, $\Omega \equiv M\!a
$, at an arbitrary time value. Calculations for suspension
d-M1-180/12 (82.5$\%$ glycerol content, $\phi_v$ = 0.016).}
\label{f:MAP}
\end{figure}

\begin{figure}[tbp]
\caption{ Dimensionless average chain length (a) and number of
aggregated particles (b) versus dimensionless rotational
frequency, $\Omega \equiv M\!a$, in a log-log plot (markers). The
power law fit (solid line) gives an exponent equal to -0.45 for
the average chain length and -1 for the $N_{a}$. Calculations for
suspension d-M1-180/12 (82.5$\%$   glycerol, $\phi_v$ = 0.016). }
\label{f:SIMULATION_RESULTS}
\end{figure}

\begin{figure}[tbp]
\caption{ Steady dichroism (a) and steady phase lag (b) as a
function of the rotating frequency for the experiments (markers)
and simulations (solid line). Measurements for suspension
d-M1-180/12 (82.5$\%$ glycerol, $\phi_v$ = 0.016). }
\label{f:H_variable_vsMA_simu}
\end{figure}


\begin{references}

\bibitem{note1} The form dichroism measures the difference between the
absorption of the incident light in the parallel and in the
perpendicular direction to the long axis of the scattering
objects. The form birefringence arises from a spatially
anisotropic arrangement of domains with different refractive
indexes.

\bibitem{Vandehulst81}  H.C. van de Hulst, Light Scattering by Small
Particles, Dover Publications, New York, 1981.

\bibitem{Melle00}  S. Melle, G.G. Fuller, and M.A. Rubio, Phys.\ Rev.\ E
{\bf 61}, 4111 (2000).

\bibitem{Helgesen90}  G. Helgesen, P. Pieranski, and A.T. Skjeltorp, Phys.
Rev. Lett. {\bf 64}, 1425 (1990); G. Helgesen, P. Pieranski, and
A.T. Skjeltorp, Phys. Rev. A {\bf 42}, 7271 (1990).

\bibitem{Meyer91}  K.B. Migler and R.B. Meyer, Phys. Rev. Lett. {\bf  66},
1485 (1991).

\bibitem{Meyer97} C. Zheng and R.B. Meyer, Phys. Rev. E {\bf 55}, 2882 (1997).

\bibitem{note2} Similar behavior was found for magnetic droplets in rotating
fields in Sandre O. et al., Phys. Rev. E {\bf 59}, 1736 (1999).

\bibitem{Gast89}  Mason number was first introduced in literature for ER
fluids under steady shear. See A.P. Gast, and C.F. Zukoski, Adv.
Colloid Interface Sci.\ {\bf 30}, 153 (1989).

\bibitem{Bossis99} O. Volkova, S. Cutillas, and G. Bossis, Phys. Rev. Lett.
{\bf 82}, 233 (1999).

\bibitem{Martin00} J.E. Martin, Phys. Rev. E {\bf 63}, 011406 (2001).

\bibitem{Jones95}  Thomas B. Jones, {\it Electromechanics of Particles},
(Cambridge, 1995).

\bibitem{Fuller95}  G.G. Fuller,{\it Optical rheometry in complex fluids},
(Oxford Universty Press, 1995).

\bibitem{note3} The values of $\lambda$ that correspond to the results reported
here are 181 $ < \lambda < $ 712 for the experiments with the
suspension d-M1-180/12 and $\lambda$= 322 for the experiments with
suspensions d-M1-070/60.

\bibitem{Martin98} J.E. Martin, R.A. Anderson, and C.P. Tigges, J.\ Chem.\ Phys.\
{\bf 108}, 3765 (1998).

\bibitem{Mohebi96}  M. Mohebi, N. Jamasbi, and Jing Liu, Phys. Rev. E {\bf 54},
5407 (1996).

\bibitem{Martin96} J.E. Martin, and R.A. Anderson, J.\ Chem.\ Phys.\
{\bf 104}, 4814 (1996).

\bibitem{private} The same behavior $L \propto M\!a^{-1/2}$ is also obtained
for a chain model in the case of rotating magnetic fields: J. E.
Martin (private communication).

\end{references}
\end{document}